\documentclass[a4paper,10pt,twoside]{cpc-hepnp}

\usepackage{multicol}
\usepackage{graphicx}
\usepackage{booktabs}
\usepackage{amssymb,bm,mathrsfs,bbm,amscd}
\usepackage[tbtags]{amsmath}
\usepackage{lastpage}
\usepackage{array}
\usepackage{supertabular}

\begin{document}

\fancyhead[c]{\small Chinese Physics C~~~Vol. XX, No. X (2015)
XXXXXX} \fancyfoot[C]{\small 010201-\thepage}

\footnotetext[0]{Received 26 April 2015}

\title{Study on CEPC performances with different collision energies and geometric layouts
\thanks{This work is supported by the Natural Science Foundation of China under the Grant No.11175192.}}

\author{%
      XIAO Ming$^{1)}$\email{xiaoming@ihep.ac.cn}
\quad GAO Jie
\quad WANG Dou \\
\quad SU Feng
\quad WANG Yi-Wei
\quad BAI Sha
\quad BIAN Tian-Jian
}
\maketitle

\address{Key Laboratory of Particle Acceleration Physics and Technology, Institute of High Energy Physics, Chinese Academy of Sciences}

\begin{abstract}
Circular Electron-Positron Collider(CEPC) is one of the largest plans in high energy physics study at China, which would serve as Higgs Factory firstly and then upgrade to a hadron collider. In this paper we give the 50km and 100km design in both single ring and double ring schemes, including $Z$ boson and $W$ boson and Higgs boson by using the optimized method. Also, we give the potential of CEPC running at $Z$ and $W$ poles. And we analysis the relationship of luminosity with circumference and filling factor, which gives a way to evaluate the choice of geometry. What's more, we compare the nominal performance of CEPC-SPPC and LHC and FCC.
\end{abstract}




\begin{multicols}{2}

\section{Introduction}

After the discovery of Higgs boson on LHC in 2012, it is natural to measure its properties as precise as possible, including mass, spin, CP nature, couplings, and etc. Compared with the International Linear Collider(ILC) working at 250GeV, a circular $e^{+}e^{-}$ collider serving as Higgs Factory seems possible due to the low mass of Higgs. And the circular scheme has the potential to upgrade to a hadron collider for high energy frontier study.
There are two ambitious international plans, one is called TLEP(renamed to FCC-ee later) at CERN aiming at constructing a 100km circular Higgs Factory, the other one is a 50km scheme starting from IHEP in Beijing.

Circular Electron-Positron Collider(CEPC) is one of the largest plans in high energy physics study at China, which would serve as Higgs Factory firstly and then upgrade to a 70-100TeV Super Proton-Proton Collider(SPPC) in the same tunnel. The goal of the CEPC is to provide $e^+e^-$ collisions at the center-of-mass energy of 240GeV where the Higgs events are produced primarily through the interaction $e^+e^-\rightarrow ZH$ and to deliver a peak luminosity greater than $1\times10^{34}\rm{cm^{-2}s^{-1}}$ per IP\cite{article1}.

$Z$ boson and $W$ boson were discovered at LEP, which have made a great contribution to particle physics. As a $e^+e^-$ collider, CEPC works as a $Z$ or $W$ Factory would be another interesting story. We use an optimized method\cite{article3} for parameter choice and compare the results of 50km scheme with 100km imagine in both single ring and double ring designs, which covers the energy region from $Z$-pole to $t$-pole. We analysis the relationship of luminosity with circumference and filling factor to evaluate the geometry choice. A comparison of nominal performance of CEPC-SPPC and LHC and FCC is also shown.

\section{Optimized method in parameter choice}

The performance of a circular $e^+e^-$ collider is connected with luminosity, which could be expressed as
\begin{equation}\label{eq:luminosity_formulae}
  L[{\rm{cm^{-2}s^{-1}}}]=2.17 \times 10^{34} (1+r) \xi_y \frac{e E_0[{\rm{GeV}}] N_b N_e}{T_{0}[{\rm{s}}] \beta^*_y[{\rm{cm}}]} F_h
\end{equation}
where $r=\frac{\sigma_y}{\sigma_x}$ is the aspect ratio of the beam at IP, $T_0$ is the revolution period, $\beta^*_y$ is the beta function at the IP, $\xi_y$ is the vertical beam-beam tune shift, $N_b$ is the bunch numbers and $N_e$ is the population of particle number in one bunch, and the $F_h$ is hour glass effect expressed as
\begin{equation}\label{eq:hourglass}
  F_h=\frac{\beta^*_y}{\sqrt{\pi} \sigma_z} {\rm{exp}}(\frac{{\beta^*_y}^2}{2{\sigma_z}^2})K_0(\frac{{\beta^*_y}^2}{2{\sigma_z}^2})
\end{equation}
where $K_0$ is the zero order modified Bessel function of the second kind.
From eq.\eqref{eq:luminosity_formulae}, it is the beam-beam tune shift that has a significant influence on the luminosity of a collider directly.

An optimized method has been well studied in \cite{article3}, which has taken several important effects into consideration, such as beam-beam limit coming from beam emittance blow-up, beam lifetime and energy spread limit constrained by beamstrahlung, and so on.
Each particle in a beam will feel a strong nonlinear force when the beam encounters the counter rotating beam, which has deleterious effects on the dynamic behavior of the particle. Within this interaction, the particles will suffer from additional heatings, which would cause beam emittance blow-up. This emittance blow-up mechanism has been studied in \cite{article2,article4}, one could get the beam-beam limit expressed as:
\begin{equation}\label{eq:beam_beam_limit}
  \xi_{y}\leq \frac{2845}{2\pi} \sqrt{\frac{T_{0}}{\tau_{y} \gamma {N_{\rm{IP}}}}}
\end{equation}
where $\rm{N_{IP}}$ is the number of interaction point, $\tau_{y}$ is the transverse damping time and $T_{0}$ is the revolution time.

Beam lifetime is determined by beamstrahlung in a high energy storage ring collider \cite{article5}. In order to achieve the beam lifetime as long as 30 minutes, one should guarantee the relationship between the bunch population and beam size satisfying
\begin{equation}\label{eq:Telnov_constraint}
  \frac{N_e}{\sigma_x \sigma_z} \leq 0.1 \eta \frac{\alpha}{3 \gamma {r_e}^2}
\end{equation}
where $N_e$ is the bunch population, $\sigma_x$ and $\sigma_z$ are the horizontal and longitudinal beam size at IP, $\alpha$ is the fine structure constant, $r_e$ is the electron classical radius and $\eta$ is the energy acceptance of ring.

Taking eq.\eqref{eq:beam_beam_limit} into eq.\eqref{eq:luminosity_formulae}, one acquires a relationship between luminosity and several key parameters of a collider
\begin{equation}\label{eq:peak_luminosity}
  L_0[{\rm{cm^{-2}s^{-1}}}]=0.7 \times 10^{34}\frac{1+r}{\beta_y^\ast [{\rm{cm}}]} \sqrt{\frac{E_0[{\rm{GeV}}] I_b[{\rm{mA}}] P_0[{\rm{MW}}]}{\gamma N_{\rm{IP}}}}
\end{equation}
\begin{equation}\label{eq:Luminosity_key_parameters}
  L[{\rm{cm^{-2}s^{-1}}}]=L_0 F_h
\end{equation}
where $E_0$ is the beam energy, $I_b$ is the average beam current, $P_0$ is the synchrotron radiation, $N_{\rm{IP}}$ is the number of interaction point and $L_0$ is the peak luminosity. From eq.\eqref{eq:peak_luminosity}, it tells us that the synchrotron radiation power is one of the pivotal parameters to the luminosity of a circular $e^+e^-$ collider. Obviously, when one tries to reduce the synchrotron radiation power, it might have deleterious effects on the luminosity.

According to the expression of $U_0=88.5\times10^3\frac{E_0^4[\rm{GeV}]}{\rho[\rm{m}]}$, there are two ordinary ways to reduce the synchrotron radiation, one is to make the machine working at lower energy, and the other one is to enlarge the bending radius. The former way leads to the plan of making CEPC severing as a $Z$ or $W$ factory, while the latter one puts forward a question: whether a 100km scheme(like FCC-ee) is better or not. Next, we will show the results by using the optimized method.

\section{Study on CEPC in different collision energies and geometric}

When restricting the synchrotron radiation power no more than 50MW, we give the parameters choice for CEPC in both 50km scheme and 100km imagine and compare the performance of double ring design with single ones. And, the potential of CEPC serving as $Z$ and $W$ factories are included. What's more, higher energy run in $t\bar{t}$ of 100km imagine are also taken into consideration. All the results are listed in TABLE \ref{tab:table1}. At this stage, we only consider all the bunches are equal around the ring and the collider is at head-on collision.

\section{Discussion}

There are many interesting topics in a circular collider ring design. Next, we will discuss three aspects about the CEPC design.

\subsection{Single ring v.s. two ring in CEPC baseline design}
Two beam pipes are adopted by many $e^+e^-$ machines, such as BEPC-II, PEP-II, KEKB and DAFNE, because high luminosity could be achieved within a large number of bunches. However, when constraining the synchrotron radiation power no more than 50MW, the average beam current is restricted at the same time because the energy loss from synchrotron radiation is the same within a certain geometry. When choosing the bunch number $N_b$ and particle population $N_e$ with a reasonable value, the luminosity of CEPC running as a Higgs factory in both one beam pipe and two are the same. Because from eq.\eqref{eq:peak_luminosity}, one could find that the luminosity is proportional to $\sqrt{P_0}$ when fixing the other parameters. It is an economic choice to take one ring scheme for a Higgs factory.

\subsection{Potential of CEPC running at $Z$ or $W$ poles}
There is an active interest in a high-luminosity run of CEPC at the $Z$ and $W$ poles. Due to lower energy of $Z$ and $W$, the synchrotron radiation in $Z$ and $W$ poles is much lower than a Higgs factory. We give the results of the parameters directly in TABLE \ref{tab:table2}. It tell us that more than 220 bunch numbers are needed in $Z$ pole to achieve luminosity as high as $1 \times 10^{34}\rm{cm^{-2}s^{-1}}$ while 60 bunches are enough for $W$ pole to reach the same luminosity. Though the synchrotron radiation power in $Z$ pole is far away from 50MW, it is beyond imagination to arrange 220 equal bunches around the ring within Pretzel Orbit. It is a vain hope to achieve a high luminosity of $1 \times 10^{35}\rm{cm^{-2}s^{-1}}$ in $Z$-pole with a 50km single ring design of CEPC within equal bunches and head-on collision, because electrostatic separators would be full of the ring to separate about 2200 bunches and the pretzel orbit would be too complicated. A bunch train scheme\cite{article6} seems hopeful to avoid this problem. However, it will make the length of the interaction regions longer and the machine-detector interface(MDI) design more complicated. So under these considerations, two beam pipes seems better.
\end{multicols}

    \begin{center}
    \setlength\tabcolsep{1pt}
    \tabcaption{ \label{tab:table1}  Comparing 50km and 100km CEPC design within single ring and double ring scheme.}
    \footnotesize
    \newcommand{\tabincell}[2]{\begin{tabular}{@{}#1@{}}#2\end{tabular}}
     \begin{tabular*}{150mm}{l|c|c|c|c|c|c|c|c|c|c|c|c|c|c}
     \hline
     \hline
     &\multicolumn{6}{c|}{50km CEPC design} &\multicolumn{8}{c}{100km CEPC design}\\
     \cline{2-15}
     \multicolumn{1}{c|}{\textbf{Parameters}} &\multicolumn{3}{c|}{Single ring scheme} &\multicolumn{3}{c|}{Double ring scheme} &\multicolumn{4}{c|}{Single ring scheme} &\multicolumn{4}{c}{Double ring scheme}\\
     \cline{2-15}
     &$Z$	&$W$ &$H$ &$Z$ &$W$ &$H$ &$Z$ &$W$ &$H$ &$t\bar{t}$ &$Z$ &$W$ &$H$ &$t\bar{t}$ \\
     \hline
       \tabincell{l}{Beam energy $E$[GeV]}	&45.5 &80 &120  &45.5 &80 &120 &45.5 &80 &120 &175 &45.5 &80 &120 &175  \\
       \hline
       \tabincell{l}{Circumference $C$[km]}	&50	&50	&50	&50	&50	&50	&100  &100  &100  &100  &100  &100  &100 &100\\
       \hline
       \tabincell{l}{Number of IP ${N_{\rm{IP}}}$} &2	&2	&2	&2	&2	&2	&2	&2	&2	&2	&2	&2	&2	&2\\
       \hline
       \tabincell{l}{Bending radius $\rho$[km]}	&6.094 &6.094 &6.094 &6.094	&6.094	&6.094	&10	&10	&10	&10	&10	&10	&10	&10\\
       \hline
       \tabincell{l}{SR power/beam $P$[MW]}	&0.89 &10.32 &50 &50 &50 &50 &0.615	&8.35 &50 &50 &50 &50 &50 &50\\
       \hline
       \tabincell{l}{SR loss/turn $U_0$[GeV]}	&0.062 &0.6 &3.01 &0.062 &0.6 &3.01	&0.038 &0.36 &1.84 &8.3 &0.038 &0.36 &1.84 &8.3\\
       \hline
       \tabincell{l}{Ring's Energy acceptance $\eta$} &0.02	&0.02 &0.02	&0.02 &0.02	&0.02 &0.02	&0.02 &0.02	&0.02 &0.02 &0.02 &0.02 &0.02\\
       \hline
       \tabincell{l}{Magnetic rigidity $B\rho$[$\rm{T\cdot m}$]} &151.8 &266.9 &400.4 &151.8 &266.9 &400.4 &151.8 &266.9 &400.4 &584 &151.8 &266.9	&400.4 &584  \\
       \hline
       \tabincell{l}{Momentum compaction \\factor $\alpha_p$[$10^{-5}$]} &0.364 &1.527 &0.729 &0.364 &1.527 &0.729 &0.453 &0.371 &0.196 &0.117 &0.453	&0.371 &0.196 &0.117 \\
       \hline
       \tabincell{l}{Lifetime due to radiative \\Bhabha scattering $\tau_L$[hour]} &8.26 &2.67 &1.19 &8.26 &2.67 &1.19 &17.6 &5.7 &2.55 &1.19 &17.6 &5.7 &2.55 &1.19\\
       \hline
       \tabincell{l}{Beam current $I$[mA]} &14.23 &16.8 &16.6 &796.81 &84.04 &16.62	&16.21 &23.02 &27.63 &5.96 &1317 &138.1 &27.63 &5.96\\
       \hline
       \tabincell{l}{Bunch number $N_b$}	&48	&48	&48	&2688 &240 &48	&192 &192 &192 &48 &15600 &1152 &192 &48\\
       \hline
       \tabincell{l}{Bunch population $N_e$[$10^{11}$]} &3.09 &3.65 &3.61 &3.09 &3.65 &3.61 &1.76 &2.5	&3.0 &2.59 &1.76 &2.5 &3.0 &2.59\\
       \hline
       \tabincell{l}{Emittance at IP-horizontal \\$\epsilon_x[\rm{nm\cdot rad}]$} &48	&18.68	&6.12 &48 &20 &6.9 &32 &18 &6.8 &2.2 &32 &18 &6.8 &2.2 \\
       \hline
       \tabincell{l}{Emittance at IP-vertical \\$\epsilon_y [\rm{pm\cdot rad}]$} &96 &36 &21.2	&96 &36 &21.2 &64 &24 &18.2 &9.2 &64 &24 &18.2 &9.2 \\
       \hline
       \tabincell{l}{Betatron function \\at IP-horizontal $\beta_x$[m]} &0.8 &0.8 &0.8 &0.8	&0.8 &0.8 &0.8 &0.8	&0.8 &0.8 &0.8 &0.8 &0.8 &0.8\\
       \hline
       \tabincell{l}{Betatron function \\at IP-vertical $\beta_y$[mm]} &1.2 &1.2 &1.2 &1.2 &1.2 &1.2 &1 &1 &1 &1 &1 &1 &1 &1   \\
       \hline
       \tabincell{l}{Transverse beam size \\at IP-horizontal $\sigma_x$[$\mu$m]} &196 &122.2	&70 &196	&122.2 &70 &160 &120 &73.8 &41.95 &160 &120 &73.8 &41.95\\
       \hline
       \tabincell{l}{Transverse beam size \\at IP-vertical $\sigma_y$[$\mu$m]} &0.339 &0.208 &0.159 &0.339 &0.208 &0.159 &0.253 &0.155 &0.135 &0.096 &0.253 &0.155 &0.135 &0.096\\
        \hline
       \tabincell{l}{Bunch length $\sigma_s$[mm]} &2.65 &2.65 &2.65 &2.65	&2.65	&2.65 &2.44 &2 &2 &1.8 &2.44 &2 &2 &1.8 \\
       \hline
       \tabincell{l}{Beam-beam parameter $\xi_x$} &0.032 &0.056 &0.112 &0.032 &0.056 &0.112 &0.028 &0.04 &0.084 &0.154 &0.028 &0.04 &0.084 &0.154\\
       \hline
       \tabincell{l}{Beam-beam parameter $\xi_y$} &0.028 &0.049 &0.074 &0.028 &0.049 &0.074 &0.022 &0.038 &0.057 &0.084 &0.022 &0.038 &0.057 &0.084  \\
       \hline
       \tabincell{l}{Hourglass factor $F_h$} &0.68 &0.68 &0.68 &0.68 &0.68 &0.68 &0.654	&0.706 &0.706 &0.732 &0.654 &0.706 &0.706 &0.732 \\
       \hline
       \tabincell{l}{Luminosity per IP \\$L$[$10^{34}\rm{cm^{-2}s^{-1}}$]} &0.22 &0.82 &1.82 &12.5 &4.08 &1.82 &0.23 &1.09 &2.93 &1.4 &18.6 &6.52 &2.93 &1.4\\
       \hline
       \tabincell{l}{RF voltage $V_{rf}$[GV]}  &0.21 &2.53 &4.98 &0.21 &2.53	&4.98 &0.36 &1.33 &2.93 &9.8 &0.36 &1.33 &2.93 &9.8 \\
       \hline
       \tabincell{l}{RF frequency $f_{rf}$[GHz]} &0.7	&1.3 &1.3 &0.7 &1.3	&1.3 &0.7 &1.3 &1.3 &1.3 &0.7 &1.3 &1.3 &1.3\\
       \hline
       \tabincell{l}{Synchrotron tune $Q_s$} &0.017 &0.127 &0.091 &0.017 &0.127 &0.09 &0.036 &0.064 &0.051 &0.049 &0.036 &0.064 &0.051 &0.049\\
       \hline
       \tabincell{l}{Energy spread $\sigma_{\delta.SR}$[\%]}  &0.05 &0.09 &0.13 &0.05 &0.09 &0.13 &0.04 &0.07 &0.10 &0.15 &0.04 &0.07 &0.10 &0.15\\
       \hline
       \tabincell{l}{Average number of photons \\emitted per electron during \\the collision $n_{\gamma}$ } &0.065 &0.122 &0.209 &0.065 &0.122 &0.209 &0.045 &0.086 &0.167 &0.253 &0.045 &0.086 &0.167 &0.253\\
       \hline
       \hline
     \end{tabular*}
    \end{center}
\begin{multicols}{2}

\subsection{Choice of the geometry}
At the moment, different geometric imagines of the future circular collider are warmly under discussion. There are two attracting plans, one is CEPC which is 50km at the preliminary design and the other one is FCC-ee which is 100km. From TABLE \ref{tab:table1}, one would find that the luminosity per IP in 100km imagine is only 1.6 times than 50km scheme. It is not economic to expense double money to gain about 60\% luminosity. However, the 100km imagine could cover the energy range of 175GeV which contains $t\bar{t}$ experiment and make it possible to upgrade to a 100Tev proton-proton collider. The advantage of a larger geometric is the possibility of higher energy frontier but not luminosity gain. So, one question comes that what size is a better choice for a Higgs factory right now?

\end{multicols}
\begin{center}
\tabcaption{ \label{tab:table2}  Parameter study for Z and W-pole under baseline design of CEPC.}
\footnotesize
\newcommand{\tabincell}[2]{\begin{tabular}{@{}#1@{}}#2\end{tabular}}
\begin{tabular*}{95mm}{@{\extracolsep{\fill}}c|c|c|c|c|c|c|c}
  \hline
  \hline
  Parameters & \multicolumn{3}{c|}{Z-pole} & \multicolumn{4}{c}{W-pole}\\
  \hline
  \tabincell{c}{$E$[GeV]} & \multicolumn{3}{c|}{45.5} & \multicolumn{4}{c}{80}\\
  \hline
  \tabincell{c}{$C$[km]} & \multicolumn{7}{c}{50}\\
  \hline
  \tabincell{c}{$N_{\rm{IP}}$}	& \multicolumn{7}{c}{2}\\
  \hline
  \tabincell{c}{$P$[MW]} &0.89 &1.85 &4.06 &10.0 &12.5 &20.8 &45.8\\
  \hline
  \tabincell{c}{$U_0$[GeV]} & \multicolumn{3}{c|}{0.62} & \multicolumn{4}{c}{0.59}\\
  \hline
  \tabincell{c}{$I$[mA]} &14.22 &29.6 &74.1 &16.8 &21.0 &35.0 &77.0\\
  \hline
  $N_b$ &48 &100 &220 &48 &60 &100 &220\\
  \hline
  \tabincell{c}{$N_e$[$10^{11}$]} & \multicolumn{3}{c|}{3.09} & \multicolumn{4}{c}{3.65}\\
  \hline
  \tabincell{c}{$\epsilon_x$[$\rm{nm\cdot rad}$]} & \multicolumn{3}{c|}{48} & \multicolumn{4}{c}{18.68}\\
  \hline
  \tabincell{c}{$\epsilon_y$[$\rm{pm\cdot rad}$]} & \multicolumn{3}{c|}{96} & \multicolumn{4}{c}{36}\\
  \hline
  \tabincell{c}{$\beta_x$[mm]} & \multicolumn{3}{c|}{0.8} & \multicolumn{4}{c}{0.8}\\
  \hline
  \tabincell{c}{$\beta_y$[mm]} & \multicolumn{3}{c|}{1.2} & \multicolumn{4}{c}{1.2}\\
  \hline
  \tabincell{c}{$\sigma_x$[m]}	& \multicolumn{3}{c|}{196} & \multicolumn{4}{c}{122.25}\\
  \hline
  \tabincell{c}{$\sigma_y$[m]}	& \multicolumn{3}{c|}{0.34} & \multicolumn{4}{c}{0.208}\\
  \hline
  \tabincell{c}{$\xi_x$} & \multicolumn{3}{c|}{0.032} & \multicolumn{4}{c}{0.056}\\
  \hline
  \tabincell{c}{$\xi_y$} & \multicolumn{3}{c|}{0.028} & \multicolumn{4}{c}{0.049}\\
  \hline
  \tabincell{c}{$\sigma_s$[mm]} & \multicolumn{3}{c|}{2.65} & \multicolumn{4}{c}{2.65}\\
  \hline
  \tabincell{c}{Hourglass factor} & \multicolumn{3}{c|}{0.68} & \multicolumn{4}{c}{0.68}\\
  \hline
  \tabincell{c}{$L$[$10^{34}\rm{cm^{-2}s^{-1}}$]} &0.22 &0.466 &1.02 &0.82 &1.02 &1.70 &3.74\\
  \hline
  \hline
\end{tabular*}
\end{center}
\begin{multicols}{2}

No matter 50km or 100km, they are general imagines for the future circular collider.It is the circumference and filling factor that affect the synchrotron radiation.
 \begin{center}
 \tabcaption{ \label{tab:table3}  Higgs Factory in different circumference.}
 \footnotesize
 \newcommand{\tabincell}[2]{\begin{tabular}{@{}#1@{}}#2\end{tabular}}
 \begin{tabular*}{75mm}{l|c|c|c}
  \hline
  \hline
  \multicolumn{4}{l}{Parameters}  \\ 
  \hline
  \tabincell{l}{Beam energy $E$[GeV]} & \multicolumn{3}{c}{120}\\
  \hline
  \tabincell{l}{Circumference $C$[km]} &50 &70 &100\\
  \hline
  \tabincell{l}{Number of IP $N_{\rm{IP}}$} & \multicolumn{3}{c}{2}\\
  \hline
  \tabincell{l}{Bending radius $\rho$[km]} &6.094 &8.60 &10.0\\
  \hline
  \tabincell{l}{SR power/beam $P$[MW]} & \multicolumn{3}{c}{50}\\
  \hline
  \tabincell{l}{SR loss/turn $U_0$[GeV]} &3.01	&2.13 &1.84 \\
  \hline
  \tabincell{l}{Beam current $I$[mA]} &16.6 &23.4 &27.6 \\
  \hline
  Bunch number $N_b$ &48 &114 &192\\
  \hline
  \tabincell{l}{Bunch population $N_e$[$10^{11}$]}	&3.61 &3.0 &3.0 \\
  \hline
  \tabincell{l}{Horizontal emittance \\$\epsilon_x$[$\rm{nm\cdot rad}$]} &6.12 &6.36 &6.8 \\
  \hline
  \tabincell{l}{Vertical emittance \\$\epsilon_y$[$\rm{pm\cdot rad}$]} &21.2 &20.0 &18.2\\
  \hline
  \tabincell{l}{Betatron function at \\IP-vertical $\beta_y$[mm]} &1.2 &1.2	&1.0\\
  \hline
  \tabincell{l}{Betatron function at \\IP-horizontal $\beta_x$[mm]}	&0.8 &0.8 &0.8\\
  \hline
  \tabincell{l}{Transverse beam size $\sigma_x$[m]}	&70.0 &71.3 &73.8\\
  \hline
  \tabincell{l}{Transverse beam size $\sigma_y$[m]}	&0.160 &0.155 &0.135\\
  \hline
  \tabincell{l}{Beam-beam parameter $\xi_x$} &0.112	&0.090 &0.084\\
  \hline
  \tabincell{l}{Beam-beam parameter $\xi_y$} &0.074	&0.062 &0.057\\
  \hline
  \tabincell{l}{Bunch length $\sigma_s$[mm]} &2.65 &2.35 &2.00\\
  \hline
  \tabincell{l}{Hourglass factor} &0.68 &0.71 &0.71\\
  \hline
  \tabincell{l}{Luminosity $L$[$10^{34}\rm{cm^{-2}s^{-1}}$]} &1.82 &2.25 &2.93 \\
  \hline
  \hline
\end{tabular*}
\end{center}
We compare the parameters in 50km, 70km and 100km, the results are shown in TABLE \ref{tab:table3}.

Using the data in TABLE \ref{tab:table3}, we give the relationship between the luminosity and circumference, which obey a power law as:
\begin{equation}\label{eq:circumference}
  L[{\rm{cm^{-2}s^{-1}}}] \sim 0.11833 \times C[{\rm{km}}]^{0.69612}
\end{equation}
And it is clearer to show it in Fig.~\ref{fig1}.
\begin{center}
\includegraphics[width=8.5cm]{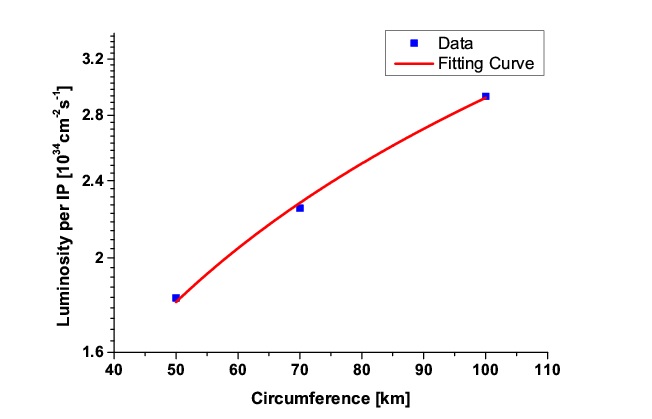}
\figcaption{\label{fig1}  Power law of Luminosity v.s. Circumference.}
\end{center}

As is known to all, the synchrotron radiation is directly connected with bending radius when the beam energy is set. Filling factor, which is defined as the length of dipoles in a ring over the circumference of the whole ring, would have an influence on the luminosity under a certain circumference. We choose 50km as example, the relationship between luminosity and filling factor is listed in TABLE \ref{tab:table4}.
\begin{table*}[htbp]
\begin{center}
\tabcaption{ \label{tab:table4}  Filling Factor.}
 \footnotesize
 \newcommand{\tabincell}[2]{\begin{tabular}{@{}#1@{}}#2\end{tabular}}
 \begin{tabular*}{92mm}{l|c|c|c|c|c|c|c}
  \hline
  \hline
  \multicolumn{4}{l}{Parameters}  \\ 
  \hline
  \tabincell{l}{Filling Factor \\$\zeta$[\%]} &70 &74 &77 &78 &80 &90 &100\\
  \hline
  \tabincell{l}{Luminosity \\$L$[$10^{34}\rm{cm^{-2}s^{-1}}$]} &1.73 &1.78 &1.82 &1.83 &1.85 &2.02 &2.07 \\
  \hline
  \hline
\end{tabular*}
\end{center}
\end{table*}
The fitting result is in eq.\eqref{eq:filling_factor}.
\begin{equation}\label{eq:filling_factor}
  L[{\rm{cm^{-2}s^{-1}}}] \sim 0.18097 \times \zeta[\%]^{0.53155}
\end{equation}
And it is clearer to show it in Fig.~\ref{fig2}.
\begin{center}
\includegraphics[width=8.5cm]{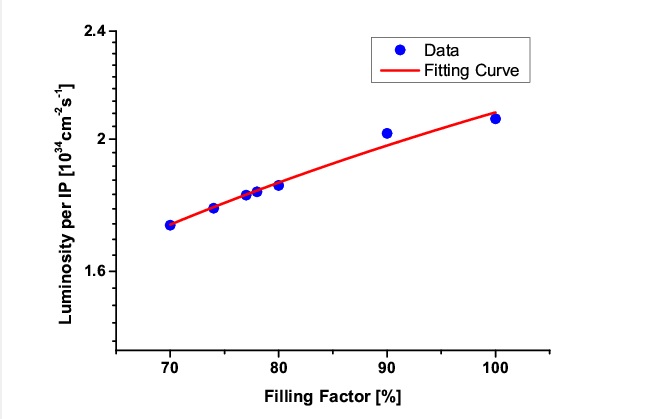}
\figcaption{\label{fig2}  Power law of Luminosity v.s. Filling Factor. The data points are from 50km design.}
\end{center}
It is the circumference and filling factor that affect the synchrotron radiation. We compare the parameters in 50km, 70km and 100km, the results are shown in TABLE \ref{tab:table3}.

In order to evaluate the geometry choice, we combine eq.\eqref{eq:circumference} and eq.\eqref{eq:filling_factor} and give the result in Fig.~\ref{fig3}. According to Fig.~\ref{fig3}, the longer in circumference and the higher in filling factor, the higher of luminosity. However, double size in circumference does not give a twice gain in luminosity from eq.\eqref{eq:circumference}, and a suitable filling factor should be taken into consideration because one should make room for other insertions around the ring. For a 50km design of circular electron positron collider, a filling factor from 60\% to 80\% is reason due to the design of other function insertions. And our choice is marked with a diamond in Fig.~\ref{fig3}.
\begin{center}
\includegraphics[width=8.5cm]{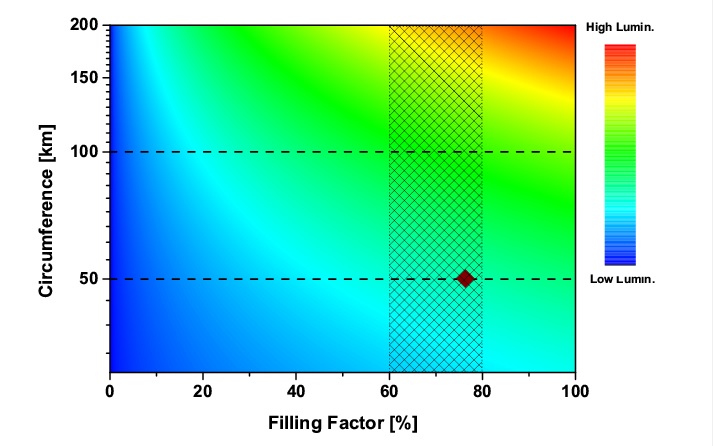}
\figcaption{\label{fig3}   The relationship of luminosity with circumference and filling factor. The shadow area shows the reasonable choice from experince. The diamond represents the choice of 50km CEPC design. }
\end{center}

Here we compare the nominal performance of CEPC-SPPC with LHC and FCC \cite{article7,article8}, and show the luminosity vs. energy in Fig.~\ref{fig4}. For CEPC and FCC-ee, the synchrotron radiation power limits the luminosity. And the expected luminosity in FCC-ee might be too high because the beam-beam parameter in \cite{article7} exceeds the beam-beam limit according to the theory in \cite{article4}. The compare results are shown in TABLE \ref{tab:table5}.
\begin{center}
\includegraphics[width=8.5cm]{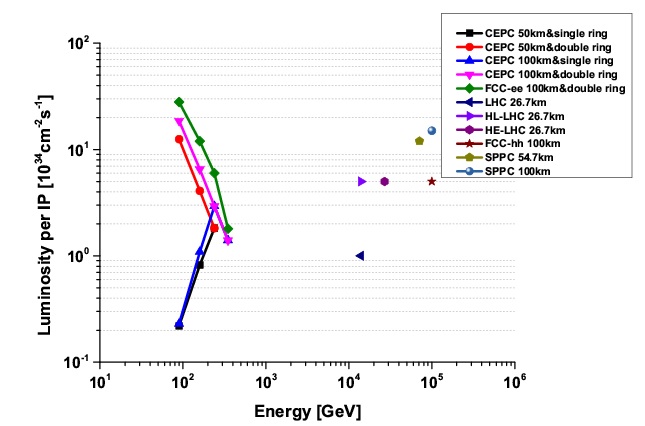}
\figcaption{\label{fig4}   Comparing the luminosity potential of CEPC-SPPC with LHC and FCC. The results are measured by the Luminosity per IP vs. Energy.}
\end{center}
\end{multicols}

\begin{table*}[htbp]
 \caption{Compare the CEPC with FCC-ee and LEP2.}
 \newcommand{\tabincell}[2]{\begin{tabular}{@{}#1@{}}#2\end{tabular}}
\begin{center}
\footnotesize
\begin{tabular}{l|c|c|c|c|c|c}
  \hline
  \hline
  Parameters & \multicolumn{1}{c|}{LEP2}  & \multicolumn{4}{c|}{FCC-ee} &CEPC   \\
  \hline
  Circumference[km] & \multicolumn{1}{c|}{26.7}  & \multicolumn{4}{c|}{100}  & 50  \\
  \hline
  Bending radius[km] & \multicolumn{1}{c|}{3.1} & \multicolumn{4}{c|}{11}   &6.094 \\
  \hline
  Momentum acceptance  &~0.01 & \multicolumn{4}{c|}{0.02}  &0.02  \\		
  \hline
  Beam energy[GeV]  &104 &45.5 &80 &120 &175 &120\\
  \hline
  IP number $N_{\rm{IP}}$ & \multicolumn{5}{c|}{4}  &2 \\
  \hline
  Beam current[mA]	 &3.04 &1450 &152 &30 &6.6 &17.45  \\
  \hline
  Bunches per beam  &4 &16700 &4490 &1360 &98 &48  \\
  \hline
  Bunch population[$10^{11}$]  &4.2 &1.8 &0.7 &0.46 &1.4 &3.79   \\
  \hline
  Transverse emittance $\epsilon$    & \multicolumn{1}{c|}{ } & \multicolumn{1}{c|}{ } & \multicolumn{1}{c|}{ } & \multicolumn{1}{c|}{ } & \multicolumn{1}{c|}{ } & \multicolumn{1}{c}{ }   \\				
  -Horizontal[nm]  &22 &29.2 &3.3 &0.94 &2 &6.9   \\
  -Vertical[pm]  &250 &60 &7 &1.9 &2 &21.2   \\
  \hline
  Momentum comp.[$10^{-5}$]  &14 &18 &2 &0.5 &0.5 &0.729  \\
  \hline
  Betatron function at IP $\beta$   & \multicolumn{1}{c|}{ } & \multicolumn{1}{c|}{ } & \multicolumn{1}{c|}{ } & \multicolumn{1}{c|}{ } & \multicolumn{1}{c|}{ } & \multicolumn{1}{c}{ }  \\
  -Horizontal[m]  &1.2 &0.5 &0.5 &0.5 &1 &0.8  \\
  -Vertical[mm]  &50 &1 &1 &1 &1 &1.2 \\
  \hline
  Beam size at IP $\sigma$[$\mu$m]  & \multicolumn{1}{c|}{ } & \multicolumn{1}{c|}{ } & \multicolumn{1}{c|}{ } & \multicolumn{1}{c|}{ } & \multicolumn{1}{c|}{ } & \multicolumn{1}{c}{ }   \\					
  -Horizontal  &182 &121 &26 &22 &45 &74.3  \\
  -Vertical  &3.2 &0.25 &0.13 &0.044 &0.045 &0.16  \\
  \hline
  Energy loss/turn[GeV]  &3.34 &0.03 &0.33 &1.67 &7.55 &3.01  \\
  \hline
  SR power/beam[MW]  &11 & \multicolumn{4}{c|}{50} &50  \\
  \hline
  Total RF voltage[GV]  &3.5 &2.5 &4 &5.5 &11 &4.98  \\
  \hline
  RF frequency[MHz] & \multicolumn{1}{c|}{352}  & \multicolumn{4}{c|}{800}  &700   \\
  \hline
  Synchrotron tune $Q_s$  &0.083 &0.65 &0.21 &0.096 &0.1 &0.09  \\
  \hline
  Hourglass factor &1 &0.64 &0.77 &0.83 &0.78  &0.68 \\
  \hline
  Luminosity/IP[$10^34\rm{cm^{-2}s^{-1}}$]  &0.012 &28 &12 &6 &1.8 &1.89  \\
  \hline
  Beam-beam parameter  & \multicolumn{1}{c|}{ } & \multicolumn{1}{c|}{ } & \multicolumn{1}{c|}{ } & \multicolumn{1}{c|}{ } & \multicolumn{1}{c|}{ } & \multicolumn{1}{c}{ }   \\				
  -Horizontal  &0.04 &0.031 &0.06 &0.093 &0.092 &0.105  \\
  -Vertical  &0.06 &0.03 &0.059 &0.093 &0.092 &0.073  \\
  \hline
  Beam-beam limit(vertical)/IP  &0.064 &0.015 &0.026 &0.038 &0.057 &0.073  \\
  \hline
  \hline
 \end{tabular}
 \end{center}
\label{tab:table5}
\end{table*}


\begin{multicols}{2}

\section{Summary}

In this paper, we give the results of CEPC performances with different collision energies and geometric layouts, including Z and W and Higgs energy run for 50km and 100km(covered $t\bar{t}$) circumference in both single ring and double ring schemes. When limiting the synchrotron radiation power to 50MW and adopting Pretzel Orbit, it is economic to construct a 50km circular electron positron collider rather than a 100km one, and one beam pipe for CEPC serving as a Higgs factory could achieve the same luminosity with double ring scheme. However, under these conditions, it is not so good to expect CEPC working in Z or W poles with high luminosity. What's more, we studied the relationship of luminosity with circumference and filling factor, which could evaluate the geometry choice. A large size of circular collider ring would be more attracting for its ability to upgrade to higher energy proton-proton collider. Also, we compare the nominal performance of the CEPC-SPPC with LHC and FCC, which shows future landscape in the high luminosity and high energy frontier.


\end{multicols}

\vspace{-1mm}
\centerline{\rule{80mm}{0.1pt}}
\vspace{2mm}

\begin{multicols}{2}

\end{multicols}

\clearpage

\end{document}